\documentclass[aps,prl,reprint,superscriptaddress,showpacs,amsmath,amssymb,floats]{revtex4-2}
\usepackage{natbib} 
\usepackage{booktabs}
\usepackage{graphicx}
\usepackage{dcolumn}
\usepackage{bm}
\usepackage[colorlinks = true,
            linkcolor = blue,
            urlcolor  = red,
            citecolor = red,
            anchorcolor = blue]{hyperref}
\usepackage{glossaries}
\newacronym{DL}{D\,--\,L}{Drude\,--\,Lorentz}
\begin{document}
\preprint{APS/123-QED}
\title{Infrared evidence for strong $C_{3}$ symmetry breaking in 1\textit{T}-TiSe$_{2}$.}

\author{Esther van Grondelle}
\email{e.r.vangrondelle@uva.nl}
\affiliation{van der Waals - Zeeman Institute, University of Amsterdam, Amsterdam, The Netherlands}
\author{Kai Rossnagel}
\affiliation{Ruprecht Haensel Laboratory, Deutsches Elektronen-Synchrotron DESY, 22607 Hamburg, Germany}
\affiliation{Institut für Experimentelle und Angewandte Physik and Ruprecht Haensel Laboratory,
Christian-Albrechts-Universität zu Kiel, 24098 Kiel, Germany}
\author{Jasper van Wezel}
\affiliation{van der Waals - Zeeman Institute, University of Amsterdam, Amsterdam, The Netherlands}
\author{Erik van Heumen}
\email{e.vanheumen@uva.nl}
\affiliation{van der Waals - Zeeman Institute, University of Amsterdam, Amsterdam, The Netherlands}

\date{\today}

\begin{abstract}
The interplay between lattice and electron degrees of freedom gives rise to competing ordered states in quantum materials, which can lead to a series of subsequent symmetry breaking transitions. While the differences are often subtle, these phases can be distinguished by their remaining point group symmetries. Here, we use infrared optical spectroscopy to probe symmetry breaking at the charge density wave transition of 1\textit{T}-TiSe$_{2}$. We uncover a previously unobserved splitting of a doubly degenerate $E_u$ optical phonon at $T_{\mathrm{CDW}}\approx$ 190 K that constitutes direct evidence for the breaking of three-fold rotational symmetry. Our finding rules out proposals where the CDW transition to a low temperature chiral or nematic phase takes place through an intermediary state that preserves three-fold symmetry. The energy difference between the two former $E_{u}$ partners can only be explained by large $C_{3}$ breaking distortions of the same order of magnitude as previously observed changes in bond lengths, suggesting that $C_{3}$ symmetry breaking plays a dominant role in the CDW transition. The linewidth of the single phonon mode above $T_{\mathrm{CDW}}$ could be consistent with a fluctuating state where three-fold symmetry is also broken. 
\end{abstract}

\maketitle

Spontaneous symmetry breaking provides a central framework for understanding phase transitions in quantum materials. The resulting electronic and structural state can be characterized through its remaining symmetries and determines the changes of its response to external probes. Quantum materials often host multiple orders that compete, coexist, or become intertwined, with their interplay giving rise to distinct patterns of symmetry breaking and emergent properties. Understanding such interrelated orders is therefore a central challenge in condensed matter physics.

1\textit{T}-TiSe$_{2}$ is a material that undergoes a phase transition to a $2a_0 \times 2a_0 \times 2c_0$ charge-density-wave (CDW) state with $T_{\mathrm{CDW}}\approx$ 189\,--\,202 K \cite{di_salvo1976,holt2001, weber2011, kogar2017}. The origin and symmetry of the CDW phase have remained elusive up to this day \cite{van_wezel2011,wegner2020,kim2024,munoz-segovia2025,wang2026,jiang2026}. The ordered state accommodates three symmetry-related CDW components, each with different orbital characteristics. Scanning tunneling microscopy (STM) experiments provided indications for rotational symmetry breaking \cite{ishioka2010} and it was proposed that charge and orbital degrees of freedom may cooperate to form a chiral charge-ordered state\,--\,an orbital-ordered state in which three-fold rotational ($C_{3}$) and inversion symmetry are broken \cite{van_wezel2011}. 

Experimental evidence for the existence of this chirality remains a topic of debate \cite{monney2010,castellan2013,hildebrand2018, lin2019,ueda2021,peng2022, kim2024,xiao2024, ueda2025}. Subsequent STM studies \cite{hildebrand2018} and x-ray diffraction (XRD) measurements \cite{ueda2021} did not find signatures of a chiral CDW phase. Thermodynamic evidence \cite{castellan2013} suggested that the chiral phase emerges via a two-step transition: first from the disordered phase to a non-chiral charge-ordered phase at $T_{\mathrm{CDW}}\approx 190 \ \text{K}$, and subsequently to a chiral charge-ordered phase at $183 \ \text{K}$. Recently, it was suggested that the chiral symmetry is already broken at $T_{\mathrm{CDW}}$ \cite{kim2024, xiao2024}. Kim \textit{et al.} report the splitting of Raman-active phonon modes at $T_{\mathrm{CDW}}$, pointing to a breaking of three-fold symmetry \cite{kim2024}. In addition, new modes observed in Raman measurements have been connected to infrared active modes. This would point to the lifting of inversion symmetry at $T_{\mathrm{CDW}}$. Electronic chirality was inferred from the observation of a strong circular dichroism \cite{xiao2024}, but it was pointed out that circular dichroism can also arise from interference effects between multi-pole scatterings \cite{ueda2025}. In addition, there have been reports that photo-induced or photo-enhanced chirality can be achieved in 1\textit{T}-TiSe$_{2}$ \cite{qiu2025,xu2020}. Finally, recent elasto-resistivity measurements have suggested that a ferroaxial order develops at $T_{\mathrm{CDW}}$ that breaks the vertical mirror planes \cite{jiang2026, edwards2026}, while transitions to $1Q$, $2Q$ or $3Q$ phases emerge for finite strain. This points to the proximity of several closely related charge ordered phases near $T_{\mathrm{CDW}}$, warranting a careful investigation of the symmetry breaking transition using a bulk sensitive probe.

In this Letter, we revisit the temperature dependence and emergence of infrared active (optical) phonons in 1\textit{T}-TiSe$_{2}$ across $T_{\mathrm{CDW}}$, recording spectra in 1 K steps. The phonon spectrum provides a bulk probe of the point group symmetry and is sensitive to the breaking of discrete symmetries. We observe a splitting of the main infrared active $E_{u}$ mode at $T_{\mathrm{CDW}}$, pointing to a lowering of the lattice symmetry and specifically to a breaking of $C_3$ symmetry. Further evidence for this comes from the distinct temperature dependent spectral weight of the two split modes compared to the modes originating from $A$, $L$ and $M$ modes that fold back onto the $\Gamma$ point. The latter show clear evidence for the charged phonon effect, the gain of phonon spectral weight due to coupling to the CDW order parameter, while the spectral weight for the former is nearly temperature independent. The breaking of $C_{3}$ symmetry rules out proposals that retain three-fold symmetry just below $T_{\mathrm{CDW}}$. 

\begin{figure}[t]
\includegraphics{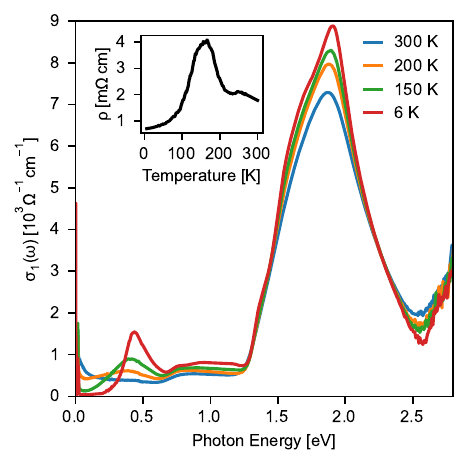} 
\caption{\label{fig:optcond} Optical conductivity, $\sigma_{1}(\omega)$, in the energy range 0-2.9 eV at selected temperatures. The inset shows the estimated optical resistivity using $\rho (T)\approx [\sigma_1(5\,\text{meV})]^{-1}$ and gives $T_{\mathrm{CDW}}\approx$ 190 K. }
\end{figure}

\textit{Infrared phonons above and below $T_{\mathrm{CDW}}$}\,--\,We employ in this study infrared Fourier-transform optical spectroscopy, a bulk-sensitive probe of the optical response. Using this technique, the reflectivity of 1\textit{T}-TiSe$_{2}$ was measured in the temperature range of 6-300 K. To obtain the full complex linear response, the Kramers-Kronig constrained Variational Dielectric Function method was used to fit the reflectivity data \cite{kuzmenko2005}. The experiment is described in more detail in appendix~\ref{app:exp}. Single crystals were grown by the iodine vapor transport method as described in the supplement of \cite{watson2019}.

The real part of the optical conductivity, $\sigma_1(\omega,T)$, is shown in Figure \ref{fig:optcond} over a wide photon energy range. Our optical data is in good overall agreement with previous results \cite{li2007, velebit2016}. For example, as previously observed by Li \textit{et al.} \cite{li2007}, 1\textit{T}-TiSe$_{2}$ has a metallic response with a low carrier density at both high and low temperatures, but with dramatically different carrier damping rates in the normal and CDW phase. Compared to Ref. \cite{velebit2016}, we extend the temperature dependence of the interband transitions in the visible range to low temperature and find that the optical response is strongly temperature-dependent even at these high energies. We do not observe the low energy mode at 60 cm$^{-1}$ of Ref. \cite{velebit2016}. 

By tracking the optical conductivity at 5 meV and assuming $\rho = 1/\sigma_1(0) \approx 1/\sigma_1(5\,\text{meV})$, we obtain an `optical' resistivity, shown in the inset of Figure \ref{fig:optcond}. 
Our data is in qualitative agreement with Refs. \cite{levy1979,rossnagel2002, moya2019, ou2024} and in quantitative agreement with Ref. \cite{moulding2022,wegner2020}. Not only is the overall temperature dependence accurately captured, but the absolute value of the resistivity is also in good agreement, confirming the reliability of our low-energy optical response measurements.

\begin{figure*}
\includegraphics[width=1\textwidth]{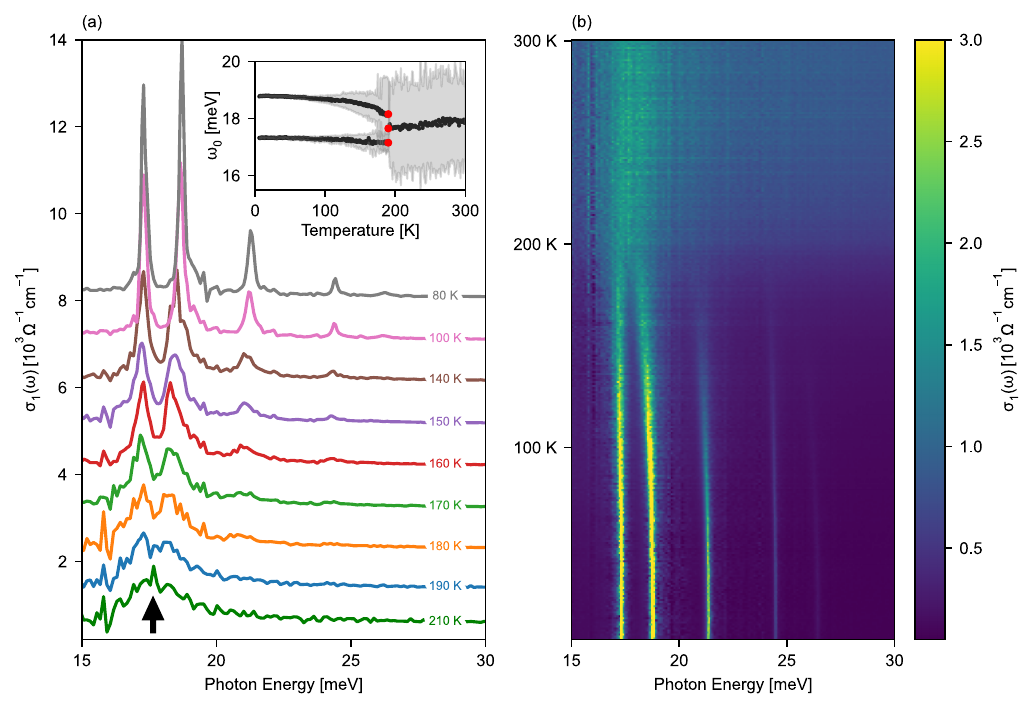} 
\caption{\label{fig:optphon}(a): Optical conductivity related to optical phonons. $\sigma_1(\omega,T)$ at each temperature is offset by $10^3\ \text{[S/cm]}$ relative to the previous $\sigma_1(\omega,T)$. The inset figure shows the eigenfrequencies of the two optical phonons obtained from fits with a Drude–Lorentz model. The gray shaded region corresponds to the phonon linewidths, with the width of the region corresponding to the linewidth $\gamma$ at that temperature.
(b) False color map of the optical conductivity showing the evolution of phonon spectra with temperature in 1 K steps. Above $T_{\mathrm{CDW}}$ a single broad phonon is visible. At the transition, two modes emerge from this single mode. At successively lower temperature, new phonon lines at 21.3, 24.5 and 26.4 meV become visible.}
\end{figure*}

One direct and in many cases unmistakable signature of a CDW phase is the appearance of new phonon modes in optical probes. These arise from the reduced lattice symmetry caused by a frozen lattice distortion. To investigate the temperature dependence of the optical phonons in the normal and CDW phase, we focus on the optical conductivity between 15 and 30 meV in Figure \ref{fig:optphon}. At high temperatures, $\text{T}>\text{T}_{\mathrm{CDW}}$  the lattice structure of 1\textit{T}-TiSe\textsubscript{2} is characterized by the $P\bar{3}m1$ (No. 164) space group, with the Ti atoms situated at the inversion centers. Given that there are three atoms per unit cell, this configuration leads to nine zone-centered vibrational modes. Decomposition of the displacements of the three atoms into irreducible representations gives the following result at the $\Gamma$ point \cite{holy1977}: 
\begin{equation}
\Gamma = A_{1g} + E_g(2) + 2 A_{2u} + 2 E_u(2).
\end{equation}
Here, the $(2)$ indicates that a mode is doubly degenerate. One A$_{2u}$ and one doubly degenerate $E_u$ mode correspond to the three acoustic branches, while the A$_{1g}$ and E$_g$ modes are Raman active. The remaining A$_{2u}$ and E$_{u}$ modes are IR active. 

Only the E$_{u}$ phonon mode is polarized in-plane and will couple to light incident perpendicular to the a-b plane \cite{holy1977,takaoka1980}. Therefore, in the normal phase, at most a single, twofold-degenerate $E_u$ optical phonon can be observed in IR optical experiments. The optical conductivity for the phonon modes of 1\textit{T}-TiSe$_{2}$ is shown in Figure \ref{fig:optphon}a at selected temperatures. The normal state $\sigma_1(\omega,T)$ at 210 K indeed shows a broad peak centered at 17.65 meV, which has been identified as the doubly degenerate $E_u$ optical phonon \cite{holy1977}. This phonon has been reported to completely soften at the $L$-point at $T_{\mathrm{CDW}}$ \cite{weber2011}. 

The phase transition at $\text{T}_{\mathrm{CDW}} \approx 190\,\text{K}$ was originally associated with a $2a_0 \times 2a_0 \times 2c_0$ charge density wave that changes the space group of the lattice to $P\bar{3}c1$ (No. 165) \cite{di_salvo1976,peng2022,holy1977}. In this superlattice structure, the unit cell contains 24 atoms instead of 3, leading to an eightfold increase in the number of phonon modes at the $\Gamma$ point as compared to the high-temperature phase. As a result, up to 10 new IR active phonon modes are expected to become observable in normal incidence reflectivity experiments \cite{holy1977}. These additional modes all originate from zone folding of the $A$, $L$, and $M$ points of the original Brillouin zone back onto the $\Gamma$ point.

Depending on the specific symmetries that are broken, zone folding and lifting of degeneracies can lead to a multitude of new phonon modes. Therefore, detailed measurement of the phonon spectrum can serve as fingerprints for the breaking of specific discrete symmetries. Experimentally, new phonon modes appear below the transition as expected. In total we observe five phonon lines at the lowest temperature of 6 K (17.3, 18.8, 21.3, 24.5 and 26.4 meV) (Fig. \ref{fig:optphon}(a,b)), which is consistent with earlier observations \cite{holy1977, li2007}. A sixth, weak phonon line can be observed in our low-temperature data at 14.6 meV. An overview of all IR and Raman phonon modes observed in low-temperature data is provided in appendix~\ref{app:modes}. The appendix also provides a list of possible mechanical representations of the phonon spectrum and the total number of expected IR and Raman active modes. It contains unit cells and space groups that derive from the high temperature $P\bar{3}m1$ space group with a three-atom unit cell and distortions previously considered in the literature. 

\textit{Evidence for lifting of degeneracy of the $E_{u}$ phonon}\,--\,The main observation in this paper is that our data shows that the doubly degenerate $E_u$ mode, indicated by an arrow in Fig. \ref{fig:optphon}(a), splits into two distinct modes at $T_{\mathrm{CDW}}$ = 190 K. Immediately below $T_{\mathrm{CDW}}$, the splitting is symmetric around the normal-phase mode with eigenfrequency 17.65 meV, which changes to two modes with eigenfrequencies 17.65 $\pm$ $\delta$, where $\delta\approx$ 0.5 meV. The development of this splitting is shown in the inset of Figure \ref{fig:optphon}(a), where the eigenfrequencies of the two optical phonons obtained from \gls*{DL} model fits are plotted as a function of temperature. The grey shaded region indicates the phonon linewidth $\gamma$ at each temperature obtained from \gls*{DL} fits. The linewidth above $T_{\mathrm{CDW}}$ is temperature independent with $\gamma_{N}\approx$ 4 meV, while the linewidths below the transition sharpen considerably. Upon further cooling, both modes sharpen and harden. The oscillator strength ($f\equiv\omega_p^2/\omega_0^2$) estimated from the \gls*{DL} models in the normal state is $f\approx$ 60, while at the lowest temperature this becomes $f_{1,2}\approx$ 22. Below we will discuss the temperature dependence of the phonon spectral weight in more detail.

The temperature dependence of the E$_{u}$ phonon contrasts with phonon modes that appear below $T_{\mathrm{CDW}}$ at higher energy, which appear without any precursor in the normal phase and are therefore likely due to zone-folding (Fig. \ref{fig:optphon}(b)). These modes gain spectral weight with decreasing temperature, while their spectral weight decreases with increasing eigenfrequency. We now discuss this in more detail.

In coupled electron-phonon systems, vibrational modes can directly gain oscillator strength from electronic spectral weight. This has become known as the charged phonon effect \cite{rice1976,rice1979,rice1992,damascelli1997,cappelluti2012}. In his original work, Rice considered the optical activity arising from the coupling between electronic degrees of freedom and the zone boundary phonons involved in the formation of the CDW. Below $T_{\mathrm{CDW}}$ the new zone-folded modes mix with phase motions of the charge density and thereby become optically active. From the perspective of lattice dynamics, two effects occur at the CDW transition. First, the distortions modify the interatomic force constants, which directly changes the phonon eigenmode frequencies and leads to mode mixing. Second, the CDW introduces a periodic modulation of bonding and anti-bonding, which creates new dipole moments in the enlarged unit cell. These dipoles couple to phonons with matching symmetry\,--\,in particular, to the zone-folded modes that involve out-of-phase motions of neighboring normal-phase unit cells\,--\,enabling them to acquire additional spectral weight (SW). In contrast, a normal-phase phonon describes largely in-phase motion, for which the CDW-induced dipoles cancel out by symmetry, leading to weak coupling. 

We track the temperature dependent spectral weight, of each individual optical phonon, $\text{SW}_{\text{phonon i}}(T) $. The phonon spectral weight is obtained by subtracting the electronic background from $\sigma_1(\omega,T)$ and integrating the optical conductivity in its corresponding energy range: $ \text{SW}_{\text{phonon i}}(T) = \int_{\omega_1}^{\omega_2} \sigma_1(\omega,T) d\omega$. This procedure can be reliably done since the background is nearly frequency independent in the relevant range (see Fig. \ref{fig:optphon}a). We define 4 frequency windows where we compute the spectral weight: Phonon 1 (15.7\,--\,17.9 meV), Phonon 2 (17.9\,--\,20.5 meV), Phonon 3 (20.8\,--\,22.3 meV) and Phonon 4 (24.1\,--\,24.9 meV). The weakest phonon at 26.4 does not have sufficient spectral weight for this analysis. 

\begin{figure}
\includegraphics{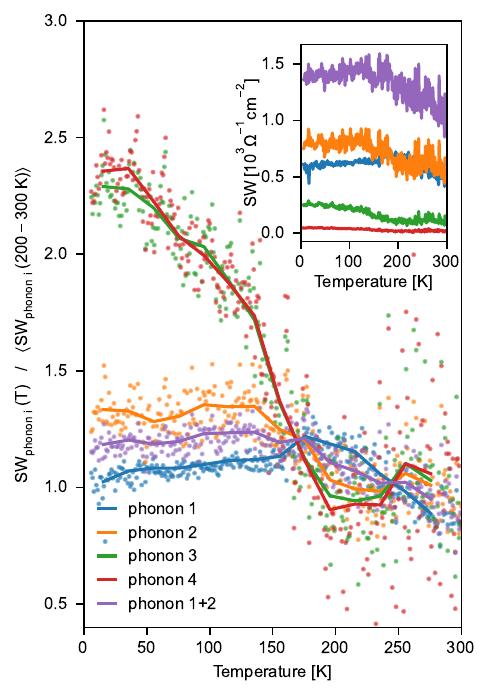}
\caption{\label{fig:SW}Spectral weight of the optical phonons as a function of temperature. The phonons are labeled in order of increasing eigenfrequency. Phonon 1 corresponds to the integrated optical conductivity between 15.7\,--\,17.9 meV after subtraction of the electronic background. Phonon 2 corresponds to 17.9\,--\,20.5 meV, phonon 3 to 20.8\,--\,22.3 meV, and phonon 4 to 24.1\,--\,24.9 meV. The line labeled phonon 1\,+\,2 corresponds to spectral weight associated with the high temperature $E_{u}$ mode and is obtained by integrating between 15.7 meV and 20.5 meV.}
\end{figure}

The total spectral weight for each of these regions is shown in the inset of Fig. \ref{fig:SW}. Above $T_{\mathrm{CDW}}$ the spectral weight of the E$_{u}$ mode is distributed over the integration windows of phonon 1, 2 and 3, due to its relatively large width. To visualize the temperature dependent phonon SW below $T_{\mathrm{CDW}}$, we use the following normalization::
\begin{equation}
 \text{SW}^{\text{Norm.}}_{\text{phonon i}}(T)= \frac{\text{SW}_{\text{phonon i}}(T)}{\langle \text{SW}_{\text{phonon i}}(200\text{-}300\,\text{K})\rangle}.
 \label{spectralweight}
 \end{equation}
 where $\left<\dots\right>$ indicates the average spectral weight in the temperature interval between 200 and 300 K. 
 
The normalized data, shown in the main panel of Fig. \ref{fig:SW}, reveals that there are two distinct temperature dependencies for the phonon spectral weights below $T_{\mathrm{CDW}}$. We can see that below $T_{\mathrm{CDW}}$ phonons 3 and 4 steadily gain spectral weight with decreasing temperature. In contrast, the spectral weight of phonon 1 and phonon 2 remain essentially independent of temperature. This distinction suggests that phonons 3 and 4 are indeed zone folded modes that derive their spectral weight from the coupling to electronic degrees of freedom, while phonons 1 and 2 are $\Gamma$-point modes that derive from the normal state $E_{u}$ mode. 

\textit{Discussion and Summary}\,--\, 
The key experimental observation presented here is that our detailed temperature dependent, bulk sensitive optical data shows that the degeneracy of the high temperature $E_{u}$ mode is lifted below $T_{\mathrm{CDW}}$. 
This implies a reduction of the point group symmetry of the unit cell relative to the normal phase. The point group of the $P\bar{3}m1$ space group associated with the lattice is $D_{3d}$ and its subgroups are $C_s$, $C_i$, $C_2$, $C_3$, $D_3$, $C_{3\nu}$, $C_{2h}$, and $S_6$.
A decomposition of the $E_u$ representation into the irreducible representations of these subgroups shows that splitting of the doubly degenerate $E_u$ mode can only occur when the symmetry is lowered to $C_s$, $C_i$, $C_2$, or $C_{2h}$. Necessarily, this symmetry reduction involves breaking the three-fold rotational symmetry ($C_3$) of the original $D_{3d}$ point group. 

Several other experiments have also pointed to breaking of $C_{3}$ symmetry. The earliest indications are from STM studies which find that the CDW amplitude modulations along the three CDW directions are unequal \cite{ishioka2010,iavarone2012}. However, STM measurements are inherently prone to surface effects, which made it desirable to obtain confirmation from bulk sensitive probes. Recent experiments that provide evidence for the development of anisotropies breaking rotational symmetry are Raman spectroscopy \cite{kim2024}, higher harmonic generation \cite{tyulnev2025}, electron diffraction \cite{wang2026} and dichroism in resonant x-ray diffraction \cite{xiao2024,xu2020}. Elasto-resistivity measurements also point to rotational symmetry breaking, albeit at finite strain or below a second transition below $T_{\mathrm{CDW}}$ \cite{jiang2026, edwards2026}. The XRD and Raman experiments also suggested the breakdown of inversion symmetry, contrasting with works reporting that inversion is retained \cite{hildebrand2018,ueda2021,jiang2026, edwards2026}. These conflicting reports could be partly related to variations in stoichiometry, disorder, or strain between samples. This seems increasingly likely given the proximity of several phases with slightly different symmetries reported in recent work \cite{jiang2026,edwards2026}. We do not believe that our observation of $C_3$-symmetry breaking at $T_{\mathrm{CDW}}$ is sample specific: the overview of previously reported phonon modes (appendix~\ref{app:modes}) are in excellent agreement with each other. In particular, the observed phonon energies would strongly depend on differences in strain.

The combination of our optical data and a recent Raman study \cite{kim2024} potentially provides insight into the breaking of inversion as well as three-fold rotations. From the Raman data, $C_3$ and inversion symmetry breaking below T$_{CDW}$ were inferred. Shifts of the $E_g$-mode energy were observed as the light polarization was rotated about the c-axis away from the crystallographic $a$-axis. This can only happen due to the lifting of the $E_{g}$ mode degeneracy and thus $C_{3}$ symmetry breaking. Inversion symmetry breaking was inferred by assigning new modes appearing below $T_{\text{CDW}}$ to infrared active $E_{u}$ modes. At low temperature, our data shows a single, well resolved infrared active mode between 16 meV and 18 meV with width $\gamma\approx$ 0.06 meV. The Raman data has four modes in this window: one $(x,y)$ pair labeled $E_{g}$ and one pair labeled $E_{u}$. 

This assignment cannot be made consistent with our data. First, the resolution and observed linewidths are identical in both experiments. The difference is therefore not due to an unresolved splitting in our experiment. In infrared spectroscopy light absorption is a first order process arising from an oscillating dipole moment. Raman scattering is a second order process and the intensity derives from the electronic polarizability \cite{larkin2017}. Because of this, a small distortion that breaks inversion symmetry could give rise to a large intensity of the $E_{u}$ mode in Raman, while the $E_{g}$ infrared spectral weight remains negligible. Therefore, the absence of $E_{g}$ modes in our infrared data does not necessarily rule out inversion symmetry breaking. 

However, the $C_{3}$ symmetry breaking lifts the $(x,y)$ degeneracy of the $E_{u}$ mode and this should show up equally in both experiments. We therefore conclude that the assignment of the $E_{u}$ mode in the Raman data is incorrect. We also note that the 18.8 meV infrared active mode, which appears below $T_{\text{CDW}}$ and which we assign to the second partner of the $E_u$ mode, is absent in the Raman spectrum. Taken together, the vibrational-mode spectra derived from both Raman and IR spectroscopy provide consistent evidence for $C_3$ symmetry breaking below $T_{\text{CDW}}$, whereas the evidence for inversion-symmetry breaking from these spectra is not conclusive. 

Inversion symmetry breaking aside, several theoretical models predict the breaking of $C_3$ or mirror symmetry in TiSe$_2$, leading to chiral \cite{van_wezel2011}, nematic \cite{munoz-segovia2025}, or ferroaxial \cite{jiang2026,edwards2026} phases. In all cases, the low-symmetry state is proposed to emerge as a small correction on top of an isotropic triple-Q CDW with symmetry  $P\bar{3}c1$ forming at $T_{\text{CDW}}$ \cite{di_salvo1976}. We can test such a scenario by estimating the magnitude of the $C_3$-breaking distortions based on the observed $E_u$ splitting. The degenerate $E_{u}$ mode involves orthogonal dipole moments along the in-plane $x$ and $y$ direction. Note that only one of these can be along the crystallographic $a$ direction and involves a single Ti-Se bond. The other mode is a linear superposition of motions involving two Ti-Se bonds. With $C_{3}$ symmetry, these modes are degenerate, implying that the corresponding average effective force constants $\langle K_x\rangle=\langle K_y\rangle\equiv \langle K\rangle$ are equal. The observed splitting requires $\langle K_x\rangle\neq\langle K_y\rangle$. Empirically, force constants relate to bond lengths through $\langle K_i\rangle\propto\langle l_i\rangle^{-n}$, where the exponent ($n\,=$ 3-7) depends on the type of bonding and whether the force constant is associated with longitudinal or transversal motion.

Since a periodic CDW distortion with a large $C_{3}$ symmetric component and a small symmetry breaking component will change average bond lengths similarly in all directions, we instead consider a uniaxial distortion that introduces a large changes in $\langle K_x\rangle$ and $\langle K_y\rangle$. Such a $C_3$-breaking distortion introduces changes in the average bond lengths along the $x$ and $y$-direction $\langle l_x\rangle\neq\langle l_y\rangle$. Using the harmonic-oscillator relation, $\omega_{0,i}=\sqrt{\langle K_{i}\rangle/m}$, and assuming that the reduced mass is unchanged, we obtain $\omega_{0,x}/\omega_{0,y}=\left(\langle l_x\rangle/\langle l_y\rangle\right)^{-n/2}$.

For the observed $E_u$ splitting at 6~K, with $\omega_{0,x}=17.3$~meV and $\omega_{0,y}=18.8$~meV, this gives $\langle l_x\rangle/\langle l_y\rangle=1.02-1.08$ for $n=3-7$. Regardless of the choice of $n$, this estimate predicts $C_{3}$ breaking distortions in the lattice of order of a few percent. For comparison, the Ti\,--\,Se bond length disproportionation obtained from low temperature neutron diffraction measurements and assigned to CDW-distortions in the isotropic triple-Q state ranges from $l_{Ti-Se1}/l_{Ti-Se2}\approx1.03$ to $l_{Ti-Se1}/l_{Ti-Se2}\approx1.07$ \cite{di_salvo1976, wegner2020}. Our estimate based on optical data is of the same order of magnitude, indicating that the $C_3$-breaking distortions cannot merely be a small perturbation on top of a much larger $C_3$-symmetric CDW distortion. Instead, it suggests that the CDW bond disproportionation is closely associated to $C_3$-breaking atomic displacements. This necessitates a theoretical approach beyond proposed chiral, nematic, or ferroaxial scenarios, which all consist of distortions on top of an isotropic triple-Q state \cite{van_wezel2011,munoz-segovia2025}.

We conclude by discussing the possible persistence of CDW fluctuations above $T_{\text{CDW}}$. Weak CDW-related band folding \cite{monney2010, ou2024,chen2016} and broad in-plane CDW Bragg peaks \cite{guo2025} have been observed to persist up to temperatures much higher than the transition temperature, suggesting that fluctuating charge order and associated symmetry reductions could be present at those elevated temperatures. 
While both infrared and Raman spectroscopy observe phonon modes consistent with the $P\bar{3}m1$ structure \cite{holy1977,li2007, velebit2016,kim2024}, we note that the high temperature phonon linewidth is much larger than that observed just below the transition. Since the phonon linewidth is sensitive to inhomogeneity, this broadening may partly arise from the spatial inhomogeneity associated with fluctuating charge order \cite{guo2025}.

It is possible that the $E_{u}$ mode with width $\gamma_{\mathrm{HT}}\approx3.1~\mathrm{meV}$ in the normal state is composed of two overlapping modes with a small splitting. Assuming a splitting $\Delta\omega$, the modes can be resolved as two peaks whenever their widths $\gamma < \sqrt{3}\Delta\omega$. At $T_{\text{CDW}}$, where we first resolve the splitting, $\Delta\omega\approx1~\mathrm{meV}$ and average width $\gamma_{avg.}\approx1.2~\mathrm{meV}$. Just above $T_{\mathrm{CDW}}$ fluctuating order will increase the linewidth of these two modes. Assuming that the splitting does not change, we can use the observed width to estimate the average $\gamma_{avg.}\approx\gamma_{obs}-(\Delta\omega)^{2}/\gamma_{obs.}$, which gives $\gamma_{avg.}\approx$ 2.8 meV. This is well above the resolution criterion $\sqrt{3}\Delta\omega\approx1.7~\mathrm{meV}$. We therefore cannot exclude that the normal-state phonon peak consists of two split modes. 

In summary, we investigated the temperature dependence and emergence of optical phonons in $1T$-TiSe$_2$ with 1 K resolution. Our high-resolution optical spectroscopy data reveals the splitting of a doubly degenerate $E_u$ mode at the CDW transition, providing direct bulk evidence for the breaking of three-fold rotational symmetry in the low-temperature phase. Combined with group-theoretical analysis, this observation rules out the previously proposed $P\bar{3}c1$ structure and constrains the symmetry of the charge-ordered state. We also argue, based on the magnitude  of the $E_u$ splitting, that the $C_3$-breaking distortion is not simply a small perturbation on top of an otherwise $C_3$-symmetric triple-Q CDW distortion, as was suggested for recently proposed chiral, nematic, and ferroaxial ground states. Consistent with the experimental results of \cite{kim2024} and \cite{xiao2024}, we moreover find no evidence for a $C_3$-breaking transition forming on top of a $C_3$-symmetric CDW through a second phase transition, as predicted in all these proposals. Instead, we observe a single CDW transition concomitant with the breaking of the three-fold symmetry for which we estimate the associated lattice distortions to be of the same order of magnitude as those associated with the CDW itself. 

\textit{Acknowledgements}\,--\, We acknowledge fruitful discussions with P. Abbamonte, P. Armitage and A. Kogar. This publication is part of the project TOPCORE (with project number OCENW.GROOT.2019.048) of the research programme Open Competition ENW Groot which is (partly) financed by the Dutch Research Council (NWO).

\bibliography{references_phonon}

\appendix
\section{Appendix A: Details of experiments and data analysis.}
\makeatletter
\edef\@currentlabel{A} 
\makeatother
\label{app:exp}

Infrared Fourier-transform optical spectroscopy is used to determine the reflectivity of 1\textit{T}-TiSe$_{2}$ between 6 and 300 K. To recover the absolute value of the reflectivity of the material as a function of temperature, the following procedure was employed: One spectrum per minute was collected while cooling the sample at a constant rate of 0.7 K per minute, followed by a similar measurement during warming up. This cycle of cooling down and warming up was repeated three times to increase the signal-to-noise ratio. By combining the data from both the warming and cooling measurements, a spectrum was obtained for every 1 K interval. To cover the full energy range, the experiments were repeated multiple times using a series of different detectors and beam splitters. Measurements in the far infrared range (3 meV\,--\,80 meV) were carried out with 0.06 meV (0.5 cm$^{-1}$) spectral resolution.

\begin{figure}
\includegraphics{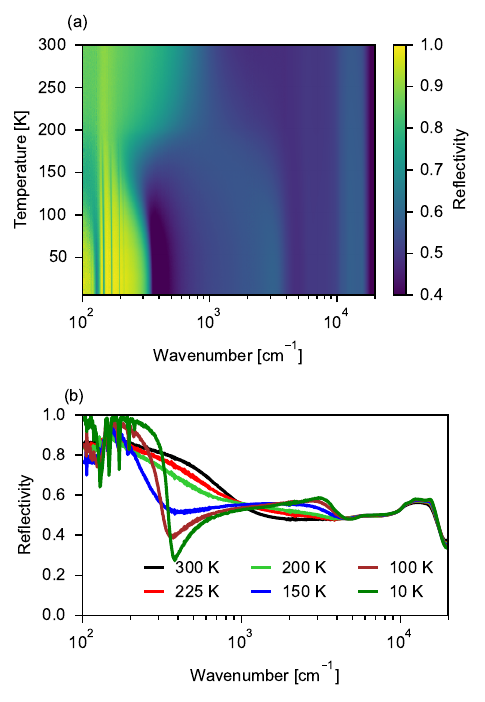}
\caption{\label{fig:reflectivity} High resolution reflectivity data for 1\textit{T}-TiSe$_{2}$ over a wide temperature range. (a): A heatmap of the reflectivity as function of temperature and wavenumber. (b): The reflectivity at selected temperatures.}
\end{figure}

To obtain an accurate absolute value of the reflectivity, metallic films were evaporated onto the sample\textit{ in situ}, and the reflectivity spectra were measured using the identical temperature cycles as the investigated sample. Reference metals with well-known, temperature-independent reflectivity spectra were used: gold for the far to mid-infrared range (3 meV\,--\,0.75 eV) and silver for the mid-infrared to visible range (0.4\,--\,2.85 eV). By comparing the ratios of the Au and Ag spectra to published literature results, we obtained standardized reference spectra as a function of temperature. These reference spectra were then used to determine the reflectivity of 1\textit{T}-TiSe$_{2}$ across the entire measured frequency range (Fig. \ref{fig:reflectivity}). For a more detailed description of the setup, please refer to Refs. \cite{tytarenko2015} and \cite{tytarenko2017}. 

To obtain the full complex linear response function the Kramers-Kronig constrained Variational Dielectric Function method is used to fit the reflectivity data \cite{kuzmenko2005}. 
This method employs a two-step fitting process. In the first step, a Drude-Lorentz model is used to fit the measured reflectivity with a minimal set of oscillators. If this model captures the major features of the experimental spectra, it can be assumed to give an approximately correct frequency dependence outside the measured spectral range, even though some fine spectral details in this energy range may not be well-fitted. In the next step, a variational dielectric function is added to describe the data. This variational function should reproduce all detailed features in the reflectivity spectrum and other optical response functions, including noise. 

\section{Appendix B: Total number of phonon modes from experiment and group theory.}
\makeatletter
\edef\@currentlabel{B} 
\makeatother
\label{app:modes}
\begin{figure}
    \centering
    \includegraphics{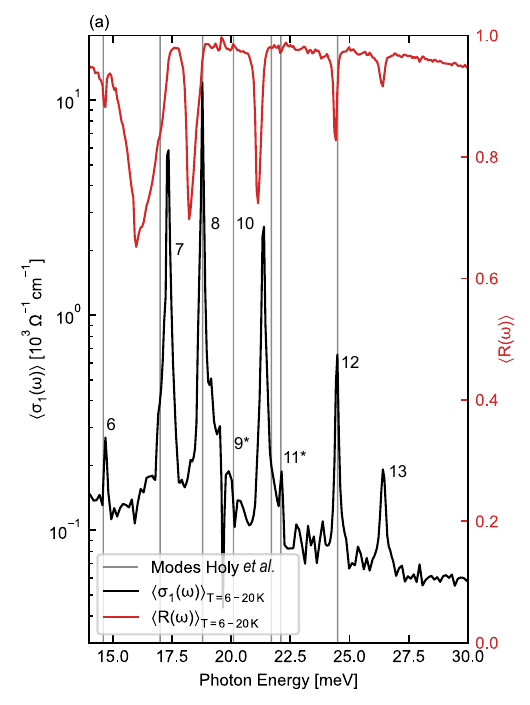}
    \caption{\label{fig:modes} (a) Low-temperature ($T =$ 6\,--\,20\,K) average reflectivity $\langle R(\omega)\rangle$ (red) and optical conductivity $\langle\sigma_1(\omega)\rangle$ (black). The grey lines indicate the phonon frequencies at 20 K reported in Ref. ~\cite{holy1977}. Modes 1\,--\,5 fall outside our experimental window; asterisk's indicate modes that are clearly resolved in our experiment, but have been reported in Ref. \cite{holy1977}. Modes 11$^{*}$ and 13 can also be observed in the reflectivity data of \cite{kim2024}.}
\end{figure}

Figure  \ref{fig:modes} compares the infrared phonon modes reported by Holy \textit{et al.}~\cite{holy1977} (gray lines) with the phonons observed in our experiment. An asterisk denotes a weak phonon mode. Holy \textit{et al.} report five infrared modes below 80 cm$^{-1}$, which are outside our detector window. As noted by the authors, the experimental reflectivity data appears not to be of sufficient quality to justify the assignment of modes below 60 cm$^{-1}$. In our combined measurements a total of 7 infrared modes are clearly resolved. In Ref. \cite{holy1977} modes 7 and 10 appear to have an asymmetry, which the authors assign to the presence of two additional modes (labeled here as 9$^{*}$ and 11$^{*}$). These modes are not resolved in our data within the noise of our experiments. However, mode 11$^{*}$ appears to be visible in the low-temperature reflectivity data of Ref. \cite{li2007} and could possibly be present in our data. Our data shows an additional mode (labeled 13) that is not present in the data of Ref. \cite{holy1977}, but is also present in the data of Ref. \cite{li2007}. Therefore, between 8\,--\,10 infrared modes have been observed. To the best of our knowledge there are no reports on the infrared active $c$-axis phonon spectrum. 

We can make a similar estimate of the total number of Raman active modes. Holy \textit{et al.} report a total 14 infrared active modes. The data of Ref. \cite{kim2024} largely reproduces this, although the data reported there only extends up to 20 meV. The improved resolution in Ref. \cite{kim2024} shows that two of the broad $E_{g}$ modes of Ref. \cite{holy1977} consists of two peaks. In addition, by rotating the light polarization about the $c$-axis it was shown that the degeneracy of the $E_{g}$ modes is lifted. This brings the total number of observed Raman active in-plane (labeled $E_{g}$) modes to be at least 7, or 14 if the degeneracy is lifted due to $C_{3}$ symmetry breaking. Together with the $A_{1g}$ modes there are then between 12 and 19 Raman active modes. 

Table \ref{tab:modes} summarizes the energies and symmetries of all experimentally observed, optically active modes. Despite the different sample sources and different experimental capabilities, the energies of the observed spectra are in very close agreement. This suggests that the expected differences in disorder, stoichiometry or distribution of CDW domain structures play a minor role and the mode spectrum is largely determined by symmetry. 

\begin{table*}
\caption{Energies of the phonon peaks observed in the normal-incidence reflectivity measured in this work, combined with those reported by \cite{holy1977} and with the Raman scattering phonon peaks from Ref. \cite{holy1977} and Ref. \cite{kim2024}. Raman modes 13\,--\,16 are from two-phonon absorption processes and should not be counted separately.}
\label{tab:modes}
\begin{tabular}{lcc lcc}
\toprule
\shortstack{Mode nr.\\ IR} & \shortstack{This work \\ $[$meV$]$}& \shortstack{Ref. \cite{holy1977} \\$[$meV$]$} &\shortstack{Mode nr.\\ Raman} & \shortstack{Ref. \cite{holy1977} \\$[$meV$]$}& \shortstack{Ref. \cite{kim2024} \\$[$meV$]$}\\
\midrule
1 & --   & 5.2 &1 ($E_g$) & -- & 8.9 \\
2 & --   & 6.4 &2 ($E_g$) & 9.2 & 9.5 \\
3 & --   & 7.9 &3 ($E_g$) & 11.5 & 11.5 \\
4 & --   & 9.4 &4 ($E_g$) & 14.1 & 13.9 \\
5 & --   & 11.2 &5 ($A_{1g}$) & 14.4 & 14.4  \\
6 & 14.6 & 14.6 &6 ($E_g$) & 16.9 & 16.9 \\
7 & 17.3 & 17.0 &7 ($E_g$) & --  & 17.1 \\
8 & 18.8 & 18.8 &8 ($E_g$) & 18.3 & --  \\
9$^*$ & -- & 20.1 &9 ($A_{1g}$) & 21.4  & --  \\
10 & 21.3 & 21.7 &10 ($A_{1g}$) & 23.2 & --  \\
11$^*$ & 22.1 & 22.1 &11 ($A_{1g}$) & 25.3  & --  \\
12 & 24.5 & 24.5 &12 ($A_{1g}$) & 37.2 & --  \\
13 & 26.4 & -- &13 ($E_g$) & 38.9 & --  \\
 & & & 14 ($A_{1g}$) & 42.8 & --  \\
  & & & 15 ($A_{1g}$) & 46.7 & --  \\
 & & & 16 ($A_{1g}$) & 50.5 & --  \\
\bottomrule
\end{tabular}
\end{table*}

Table \ref{tab:symmetry-modes} summarizes the expected number of in-plane infrared and Raman active phonon modes for a number of space groups and structures. To arrive at this list, we considered unit cell doublings and space groups that fit previously reported symmetry breakings at $T_{\mathrm{CDW}}$, starting from the high temperature $P\bar{3}m1$ structure with a unit cell containing one Ti and two Se atoms. We considered reconstructions where the unit cell doubles in one in-plane direction and the out-of plane direction (e.g. 2$\times$1$\times$2), or in two in-plane directions and the out of plane direction (e.g. 2$\times$2$\times$2). In general, the in-plane lattice constant can change (involving a $Q$\,=\,0 distortion) and for some space groups the $a$ and $b$ axes will become different. We subsequently used the Bilbao crystallographic server to deduce the full mechanical representation and from that the optically active in-plane modes \cite{kroumova2003}. 

The $2a\times1b\times2c$ unit cell explicitly distinguishes the $a$ and $b$ directions, lifting the degeneracy of the $E$ modes for all space groups considered in this category. For the $2\times2\times2$ CDW structure, we include both space groups proposed in the literature that preserve and those that break threefold rotational symmetry. While this list is not exhaustive, it provides an important check on proposed space groups for the low-temperature phase.

\begin{table*}[t]
\caption{Overview of space groups for different super cells. Second column indicates the number of atoms in the primitive cell, which is necessary to correctly count the total degrees of freedom. Third column gives the full mechanical representation of the acoustic, silent, infrared and Raman modes. Fourth and fifth column give the expected number of infrared and Raman active modes respectively. The final column indicates references where the structure is discussed or considered.}
\label{tab:symmetry-modes}
\begin{tabular}{lc lll c}
\toprule
Structure & \shortstack{Atoms in the \\primitive u.c.} & Mechanical Representation & \shortstack{IR\\(in-plane)} & \shortstack{Raman\\active} & Refs. \\
\midrule
$a_{0}\times a_{0}\times c_{0}$\\
$P\bar{3}m1$ & 3 & $\Gamma = A_{1g} + 2A_{2u} + E_g(2) + 2E_u(2)$ & $E_u(2)$ & $E_g(2)$ & \\
$2a\times b\times2c$\\
$C2/m$ & 6 & $\Gamma = 4A_u + 4A_g + 8B_u + 2B_g$ & $6B_u$ & $4A_g + 2B_g$ & \\
$C2/c$ & 6 & $\Gamma = 4A_u + 4A_g + 5B_u + 5B_g$ & $3B_u$ & $4A_g +5B_g$ & \\
$Cm$ & 6 & $\Gamma = 10A^{\prime} + 8A^{\prime\prime}$ & $8A^{\prime}$ & $8A^{\prime}+ 7A^{\prime\prime}$ & \\
$Cc$ & 6 & $\Gamma = 9A^{\prime} + 9A^{\prime\prime}$ & $7A^{\prime}$ & $7A^{\prime}+ 8A^{\prime\prime}$ & \\
$C2$ & 6 & $\Gamma = 8A + 10B$ & $8B$ & $7A +8B$ & \\
$P2/c$ & 12 & $\Gamma = 12A_u + 6A_g + 12B_u + 6B_g$ & $10B_u$ & $6A_g + 6B_g$ & \\
$P\bar{1}$ & 12 & $\Gamma = 15A_g + 21A_u$ & $18A_u$ & $15A_g$ & \\
$P1$ & 12 & $\Gamma = 36A$ & $33A$ & $33A$ & \\
$2a\times2a\times2c$\\
$P\bar{3}m1$ & 24 & $\Gamma = 9A_{1g} + 3A_{2g} + 3A_{1u} + 9A_{2u} + 12E_g(2) + 12E_u(2)$ & $11E_u(2)$ & $9A_{1g} +12E_g(2)$ & \cite{wegner2020}\\
$P\bar{3}c1$ & 24 & $\Gamma = 5A_{1g} + 7A_{1g} + 5A_{1u} + 7A_{2u} + 12E_g(2) + 12E_u(2)$ & $11E_u(2)$ & $7A_{1g} +12E_g(2)$ & \cite{holy1977}\\
$P321$ & 24 & $\Gamma = 14A_2 + 10A_1 + 24E(2)$ & $23E(2)$ & $10A_1 + 23E_g(2)$ & \cite{kim2024}\\
$P\bar{3}$ & 24 & $\Gamma = 8A_g + 16A_u + 8E_g(2) + 16E_u(2)$ & $15E_u(2)$ & $8A_g +8E_g(2)$ & \cite{jiang2026,edwards2026}\\
$2a\times2b\times2c$\\
$C2/m$ & 12 & $\Gamma = 8A_u + 9A_g + 13B_u + 6B_g$ & $11B_u$ & $9A_g +6B_g$ & \\
$C2/c$ & 12 & $\Gamma = 10A_u + 7A_g + 11B_u + 8B_g$ & $9B_u$ & $7A_g +8B_g$ & \cite{subedi2022}\\
$C2$ & 12 & $\Gamma = 17A+19B$ & $17B$ & $17A+17B$ & \cite{xu2020,subedi2022}\\
$Cm$ & 12 & $\Gamma = 21A^{\prime} + 15A^{\prime\prime}$ & $19A^{\prime}$ & $19A^{\prime}+ 14A^{\prime\prime}$ & \\
$Cc$ & 12 & $\Gamma = 18A^{\prime} + 18A^{\prime\prime}$ & $16A^{\prime}$ & $16A^{\prime}+ 17A^{\prime\prime}$ & \cite{subedi2022}\\
$P2/c$ & 24 & $\Gamma = 24A_u + 12A_g + 24B_u + 12B_g$ & $22B_u$ & $12A_g +12B_g$ & \cite{feya2021}\\
$P2$ & 24 & $\Gamma = 34A+38B$ & $36B$ & $34A+36B$ & \cite{van_wezel2011}\\
$P\bar{1}$ &24 & $\Gamma = 30A_g + 42A_u$ & $39A_u$ & $30A_g$ & \\
$P1$ & 24 & $\Gamma = 72A$ & $69A$ & $69A$ & \cite{kim2024}\\
\bottomrule
\end{tabular}
\end{table*}

\end{document}